\documentclass[10pt]{article}
\usepackage[letterpaper,top=2cm,bottom=2cm]{geometry}
\usepackage{graphicx}
\usepackage{hyperref}
\usepackage{amssymb}
\usepackage{amsmath}
\usepackage{slashed}
\usepackage[dvipsnames]{xcolor}
\usepackage[utf8]{inputenc}
\usepackage[style=phys]{biblatex}
\usepackage{mathtools}
\usepackage{relsize}

\newcommand\pT{\ensuremath{p_\mathrm{T}}}

\newcommand\pubdate{\today}

\def\Title#1{\begin{center} {\Large #1 } \end{center}}
\def\Author#1{\begin{center}{ \sc #1} \end{center}}
\def\Address#1{\begin{center}{ \it #1} \end{center}}

\newcommand\pubblock{\rightline{\begin{tabular}{l}  \\ 
         \pubdate  \end{tabular}}}
\newenvironment{Abstract}{\begin{quotation}  }{\end{quotation}}
\newenvironment{Presented}{\begin{quotation} \begin{center} 
             PRESENTED AT\end{center}\bigskip 
      \begin{center}\begin{large}}{\end{large}\end{center} \end{quotation}}

\begin{document}
\DeclarePairedDelimiter\abs{\lvert}{\rvert}

\begin{titlepage}
 \pubblock
\vfill
\Title{Measurement of CollinearDrop jet mass and its correlation with SoftDrop groomed jet substructure observables in $\sqrt{s}=200$ GeV $pp$ collisions by STAR}
\vfill
\Author{ Youqi Song (Wright Laboratory, Yale University)}
\Address{on behalf of the STAR Collaboration}
\vfill
\begin{Abstract}
Jet substructure variables aim to reveal details of the parton fragmentation and hadronization processes that create a jet. By removing collinear radiation while maintaining the soft radiation components, one can construct CollinearDrop jet observables, which have enhanced sensitivity to the soft phase space within jets. We present a CollinearDrop jet measurement, corrected for detector effects with a machine learning method, MultiFold, and its correlation with groomed jet observables, in $pp$ collisions at $\sqrt{s}=200$ GeV at STAR. We demonstrate that the population of jets with a large non-perturbative contribution can be significantly enhanced by selecting on higher CollinearDrop jet mass fractions. In addition, we observe an anti-correlation between the amount of grooming and the angular scale of the first hard splitting of the jet.
\end{Abstract}
\vfill
\begin{Presented}
DIS2023: XXX International Workshop on Deep-Inelastic Scattering and
Related Subjects, \\
Michigan State University, USA, 27-31 March 2023
\end{Presented}
\vfill
\end{titlepage}

\section{Introduction}

Jets are collimated sprays of final-state hadrons produced from initial high-momentum transfer partonic scatterings in particle collisions. One class of jet clustering algorithms, sequential recombination algorithms, relies on recursively recombining two particles/objects into one object by minimizing a distance metric between the two, mimicking the QCD parton shower pattern \cite{Marzani_2019}. Since jets are multi-scale objects that connect asymptotically free partons to confined hadrons, jet substructure measurements can provide insight into the parton evolution and the ensuing hadronization processes. 

To enhance perturbative contributions, SoftDrop grooming \cite{larkoski2014soft} is often used to remove wide-angle soft radiation from jets. The procedure, detailed in Ref. \cite{larkoski2014soft}, starts by re-clustering the jet with an angular-ordered sequential recombination algorithm called Cambridge/Aachen \cite{Dokshitzer:1997in, Wobisch:1998wt}. Then the last step of the clustering is undone and the softer prong of the two is removed until the SoftDrop condition specified by $(z_{\mathrm{cut}}, \beta)$ is met:

\begin{equation}
    \label{eq:zg}
    z_\mathrm{g}=\frac{\mathrm{min}(p_{\mathrm{T,1}},p_{\mathrm{T,2}})}{p_{\mathrm{T,1}}+p_{\mathrm{T,2}}}>z_\mathrm{cut}(R_\mathrm{g}/R_\mathrm{jet})^{\beta},
\end{equation}
where $z_\mathrm{cut}$ is the SoftDrop momentum fraction threshold, $\beta$ is an angular exponent, $R_\mathrm{jet}$ is the jet resolution parameter, $p_{\mathrm{T,1,2}}$ are the transverse momenta of the two subjets, and $R_{\mathrm{g}}$ is defined as:

\begin{equation}
    \label{eq:rg}
    R_\mathrm{g} = \sqrt{(y_1-y_2)^2 + (\phi_1-\phi_2)^2},
\end{equation}
where $y_{1,2}$ and $\phi_{1,2}$ are, respectively, the rapidity values and azimuthal angles of the two subjets. $z_\mathrm{g}$ and $R_\mathrm{g}$ describe the momentum imbalance and the opening angle of the SoftDrop groomed jet, respectively. 

Although the SoftDrop groomed jet substructure observables have been extensively studied both experimentally \cite{CMS:2017qlm, Aad:2705512, adam2020measurement, ATLAS:2017zda, CMS:2018fof, abdallah2021invariant} and theoretically \cite{Larkoski:2017jix}, the effects of wide-angle and soft radiation which are suppressed by SoftDrop measurements, have not yet been explored as much. 
One set of observables that are sensitive to the soft wide-angle radiation are known as the CollinearDrop jet observables \cite{chien2020collinear}. The general case involves the difference of two different SoftDrop selections $(z_{\mathrm{cut},1}, \beta_1)$ and $(z_{\mathrm{cut},2}, \beta_2)$, which reduces the collinear contributions from fragmentation, and the wide-angle contributions from initial-state radiation (ISR), underlying event (UE) and pileup. For this analysis, the less aggressive SoftDrop grooming criteria is set to no grooming, $(z_{\mathrm{cut},1}, \beta_1)=(0,0)$, so the CollinearDrop groomed observables are the difference in the ungroomed and SoftDrop groomed observables. This simplification can be made since the wide-angle contributions from ISR, UE and pileup are not significant for the dataset used in this analysis. Specifically, the contribution of UE to jet \pT\ for a jet with $20 < \pT < 25$ GeV$/c$ is less than 1\% \cite{adam2020underlying}. 

As the QCD parton shower is angular ordered \cite{Dokshitzer:1991wu}, the soft wide-angle radiation captured by the CollinearDrop jet observables on average happens at an early stage of the shower. Unlike the opposite approach of CollinearDrop, SoftDrop will then capture the late stage collinear splittings. Therefore, a simultaneous measurement of a CollinearDrop jet observable and a SoftDrop jet observable can help illustrate the dynamics of the parton shower. 

The CollinearDrop jet observable that we focus on is $\Delta M$ or $\Delta M/M$, defined as:

\begin{equation}
    \Delta M/M = \frac{M-M_\mathrm{g}}{M},
\end{equation}
where $M$ is the jet mass and $M_{\mathrm{g}}$ is the SoftDrop groomed jet mass, defined as:

\begin{equation}
    \label{eq:m}
    M_{(\mathrm{g})} = \abs*{{\mathlarger\sum}_{i \in \mathrm{(groomed)\ jet}} p_i} = \sqrt{E_{(\mathrm{g})}^2-|\vec{\mathbf{p}}_{(\mathrm{g})}|^2 },
\end{equation}
where $p_i$ is the four-momentum of the $i$th constituent in a (groomed) jet, and $E_{(\mathrm{g})}$ and $\vec{\mathbf{p}}_{(\mathrm{g})}$ are the energy and three-momentum vector of the (groomed) jet, respectively.

The CollinearDrop groomed mass is also expressed as \cite{chien2020collinear}:

\begin{equation}
    a = \frac{M^2-M_{\mathrm{g}}^2}{p_{\mathrm{T}}^2}.   
\end{equation}
where $p_{\mathrm{T}}^2$ is the jet transverse momentum squared. While $\Delta M$ is a more straightforward measure of the amount of wide-angle soft radiation, $a$ is calculable in Soft Collinear Effective Field Theory (SCET) at the parton level \cite{chien2020collinear}.

In these proceedings, we present measurements of the CollinearDrop groomed jet mass, to study the less-explored phase space of soft and wide-angle radiation; we also measure the correlation between the CollinearDrop groomed mass with $R_{\mathrm{g}}$ and $z_{\mathrm{g}}$, in $pp$ collisions at $\sqrt{s}=200$ GeV at STAR. The measurements are fully corrected for detector effects with MultiFold, a novel machine learning method which preserves the correlations in the multi-dimensional observable phase space. We then compare our fully corrected measurements with predictions from event generators.

\section{Analysis details}
\label{analysis_details}
The STAR experiment \cite{STAR:2002eio} recorded data from $\sqrt{s} = 200$ GeV $pp$ collisions during the 2012 RHIC run. Tracks are reconstructed from the Time Projection Chamber (TPC), and neutral energy deposits are measured from the Barrel Electro-Magnetic Calorimeter (BEMC) towers. Events are selected within $\pm 30$ cm from the center of the detector along the beam axis, and to pass the jet patch trigger which requires a minimum transverse energy $E_{\mathrm{T}}>7.3$ GeV deposited in a $1 \times 1$ patch in $\eta \times \phi$ in the BEMC. The same track and tower selections are applied as in Ref. \cite{adam2020measurement} and \cite{abdallah2021invariant}. We reconstruct jets from TPC tracks ($0.2 < \pT < 30\ \mathrm{GeV}/c$, with a charged pion mass assignment) and BEMC towers ($0.2 < E_{\mathrm{T}} < 30\ \mathrm{GeV}$, assuming massless) using the anti-$k_{\mathrm{T}}$ sequential recombination clustering algorithm \cite{cacciari2008anti} with a resolution parameter of $R=0.4$. We apply the selections of $p_{\mathrm{T}} > 15$ GeV$/c$, $|\eta|<0.6$, transverse energy fraction of all neutral components $<0.9$, and $M>1$ GeV$/c^2$ on reconstructed jets, consistent with the selections in Ref. \cite{abdallah2021invariant}. We select jets that pass SoftDrop grooming with the standard cuts of $(z_{\mathrm{cut}}, \beta)=(z_{\mathrm{cut},2}, \beta_2)=(0.1,0)$.

We measure the following jet observables: $p_\mathrm{T}$, $z_{\rm{g}}$ (defined in Eq. \ref{eq:zg}), $R_{\rm{g}}$ (defined in Eq. \ref{eq:rg}), $M$ (defined in Eq. \ref{eq:m}), $M_{\rm{g}}$ (defined in Eq. \ref{eq:m}), and jet charge $Q^{\kappa=2}$. $Q^{\kappa=2}$ is defined as:

\begin{equation}
    Q^{\kappa=2}=\frac{1}{p_{\mathrm{Tjet}}^2}\mathlarger{\sum}_{i \in \mathrm{jet}} \ q_i \cdot p_{\mathrm{T}_i}^2,
\end{equation}
where $q_i$ and $p_{\mathrm{T}_i}$ are the electric charge and \pT\ of the $i$th jet constituent, respectively.

Experimentally, jet measurements need to be corrected for detector effects to compare with theoretical calculations and model predictions. The traditional correction procedure uses the Bayesian inference in as many as three dimensions and requires the observables to be binned \cite{d2010improved}. A novel correction procedure, MultiFold \cite{andreassen2020omnifold}, uses a machine learning technique to correct with higher dimensionality in an un-binned fashion. Because it preserves the correlation between observables with high dimensionality, MultiFold is potentially more desirable for this study.

We fully corrected these six jet observables simultaneously for detector effects using MultiFold. The particle-level prior used for unfolding is jets from events generated with PYTHIA6 \cite{Sjostrand:2006za} with the STAR tune \cite{Adkins:2015ccl}. This is a single-parameter modification to the Perugia 2012 tune \cite{Skands:2010ak} to better match STAR data. The PYTHIA events are run through GEANT3 \cite{Brun:1994aa} simulation of the STAR detector, and embedded into data from zero-bias events from the same run period as the analyzed data. The detector-level prior is the jets reconstructed after this embedding procedure and geometrically matched to the particle-level truth jets. The correction was validated using a Monte Carlo closure test, which showed good performance of the unfolding among all observables for jets with $20<p_{\mathrm{T}}<50$ GeV$/c$. To correct for fake jets, i.e., detector-level jets arising from background, fake rates were obtained from simulations and used as initial weights for the data as an input to MultiFold. For particle-level jets that are missed at detector level due to effects such as tracking inefficiency, an efficiency correction was done post-unfolding in a multi-dimensional fashion.

\begin{figure}[ht!]
    \centering
    \includegraphics[width=0.98\textwidth]{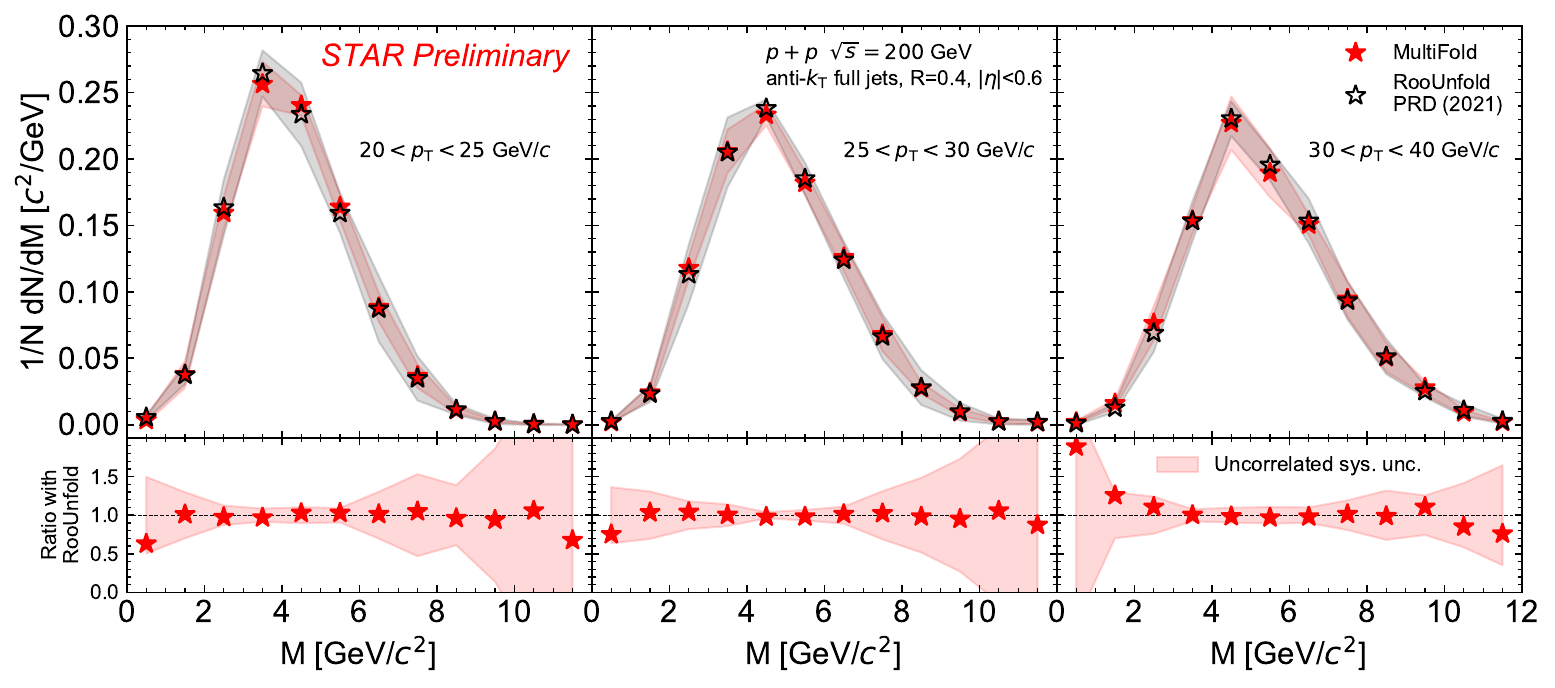}
    \caption{Jet mass distributions corrected with MultiFold and RooUnfold.}
    \label{fig:m}
\end{figure}

The fully corrected jet mass distributions for three different \pT\ bins are shown in Fig. \ref{fig:m}, using both MultiFold and RooUnfold \cite{abdallah2021invariant}. In the top panels, the distributions obtained with MultiFold are shown to agree with the previously published RooUnfold result; in the bottom panels, the ratios of MultiFold distributions over RooUnfold distributions are confirmed to be consistent with unity. These establish further confidence in application of MultiFold to the data.

The dominant source for systematic uncertainty is from unfolding prior variation. For the MultiFold measurement in Fig. \ref{fig:m}, it is accounted for by varying the prior shape of \pT\ and $M$, as in Ref. \cite{abdallah2021invariant}. For the following MultiFold measurements, it is accounted for through simultaneous reweighting of all six observables based on prior distributions from PYTHIA \cite{aguilar2022pythia} and HERWIG \cite{bellm2016herwig}.

\section{Results}
\begin{figure}[ht!]
    \centering
    \includegraphics[width=0.48\textwidth]{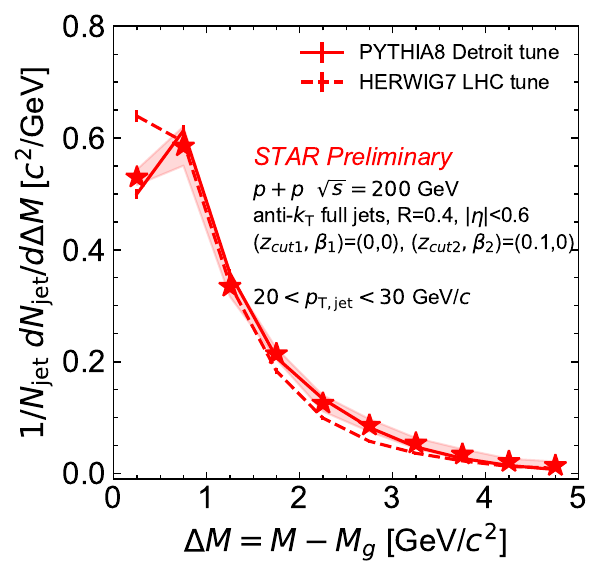}
    \includegraphics[width=0.48\textwidth]{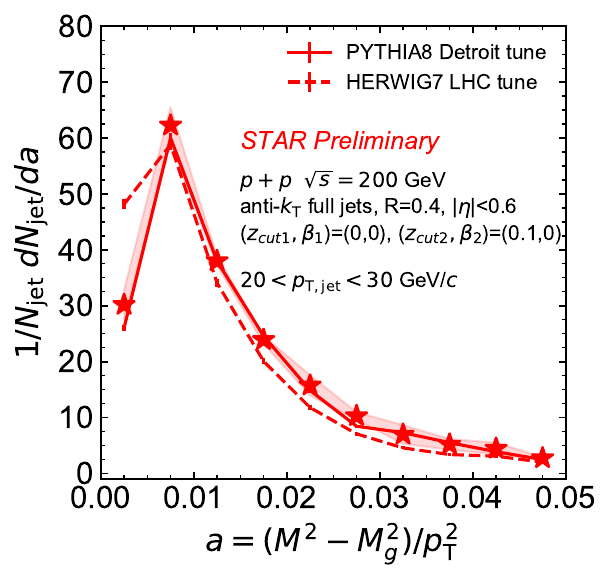}
    \caption{CollinearDrop jet mass distributions. The error bands indicate systematic uncertainties.}
    \label{fig:dm}
\end{figure}

Figure \ref{fig:dm} shows the distribution of fully corrected CollinearDrop groomed jet masses for jets within $20 < \pT < 30$ GeV$/c$. This measurement excludes jets with $M = M_\mathrm{g}$ (46\% of jets in this \pT\ range) so that the peak in the small but nonzero $\Delta M$ region is visible. Both PYTHIA8 Detroit tune (tuned to RHIC kinematics) \cite{aguilar2022pythia} and HERWIG7 H7.1-Default
tune (tuned to LHC kinematics) \cite{bellm2016herwig} capture the qualitative trend of data, although there is some tension with HERWIG in the small $\Delta M$ region.

\begin{figure}[ht!]
    \centering
    \includegraphics[width=0.525\textwidth]{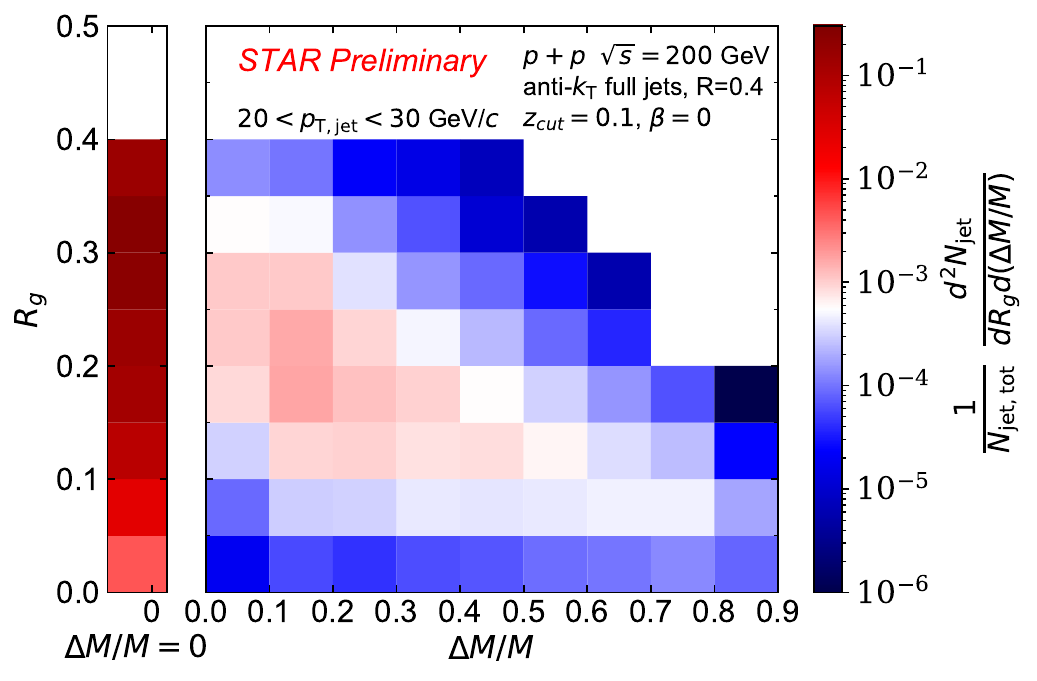}
    \includegraphics[width=0.425\textwidth]{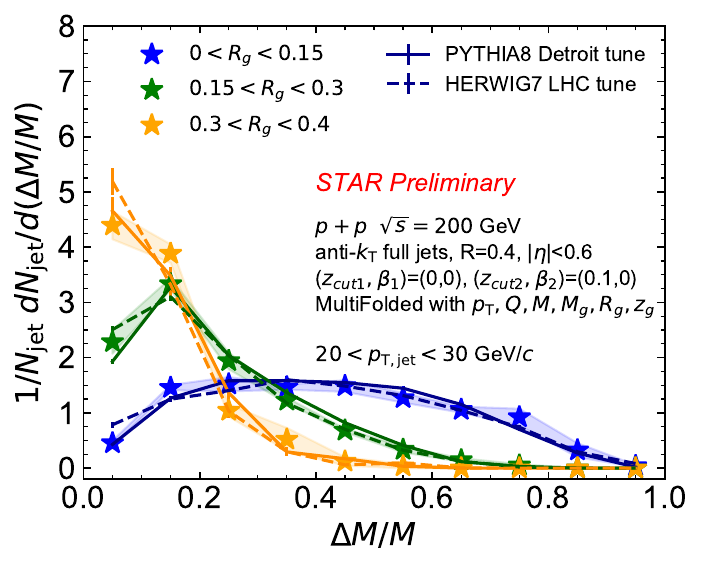}
    \caption{Correlation between the CollinearDrop groomed mass fraction $\Delta M/M$ and the SoftDrop groomed jet opening angle $R_{\mathrm{g}}$.}
    \label{fig:dmfrac_rg}
\end{figure}

Figure \ref{fig:dmfrac_rg} (left) shows the correlation between the CollinearDrop groomed mass fraction $\Delta M/M$ and the SoftDrop groomed jet opening angle $R_{\mathrm{g}}$. The jet population with $M = M_{\mathrm{g}}$ has been separated out into the leftmost column. For the $M > M_{\mathrm{g}}$ panel, a diagonal trend that indicates an anti-correlation between the amount of soft radiation and the hard splitting angle is observed, consistent with the expectation of angular ordering of the parton shower. Figure \ref{fig:dmfrac_rg} (right) shows the projection of $\Delta M/M$ for different selections of $R_{\mathrm{g}}$. We observe that, as shown in the blue data points, a selection on small $R_{\mathrm{g}}$ results in a relatively wide $\Delta M/M$, suggesting that a small SoftDrop groomed jet radius appears with a wide range of SoftDrop grooming. On the other hand, as shown in the orange data points, a selection on large $R_{\mathrm{g}}$ results in a sharper $\Delta M/M$ peaked towards small $\Delta M/M$ values, suggesting that a large SoftDrop groomed jet radius leaves space for little or no SoftDrop grooming. This measurement demonstrates how early soft wide-angle radiation constrains the angular phase space of later splittings. PYTHIA and HERWIG predictions, as indicated by the solid and dashed lines, describe the trends of the data.

\begin{figure}[ht!]
    \centering
    \includegraphics[width=0.425\textwidth]{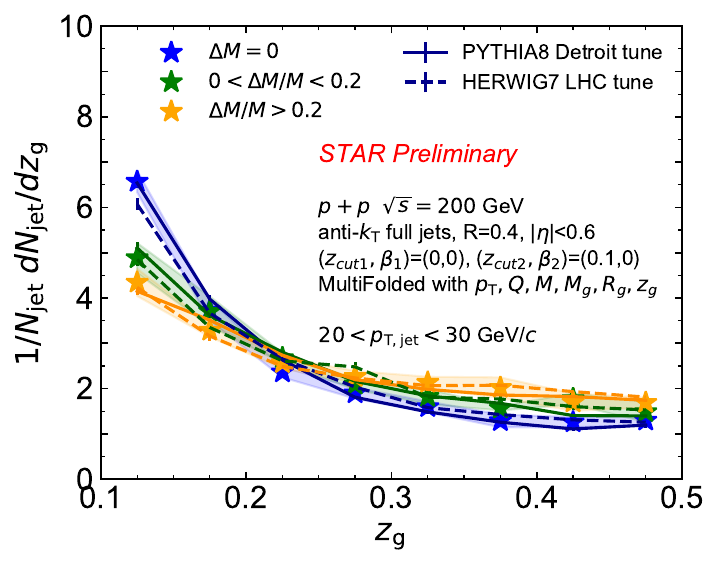}
    \caption{Correlation between the CollinearDrop groomed mass fraction $\Delta M/M$ and the SoftDrop groomed jet opening angle $z_{\mathrm{g}}$.}
    \label{fig:zg_dmfrac}
\end{figure}

Figure \ref{fig:zg_dmfrac} shows the correlation between $\Delta M/M$ and the SoftDrop groomed shared momentum fraction $z_{\mathrm{g}}$. We observe that the more fractional mass that is groomed away by SoftDrop, the flatter the $z_{\mathrm{g}}$ distribution is. Since the perturbative DGLAP splitting function follows the $1/z$ behavior \cite{Gribov:1972ri, ALTARELLI1977298, Dokshitzer:1977sg}, a more steeply falling $z_{\mathrm{g}}$ with small $\Delta M$ indicates a larger perturbative contribution. This measurement demonstrates how an early-stage emission constrains the momentum imbalance of a later splitting. Again, PYTHIA and HERWIG are able to describe the data.

\begin{figure}[ht!]
    \centering
    \includegraphics[width=0.99\textwidth]{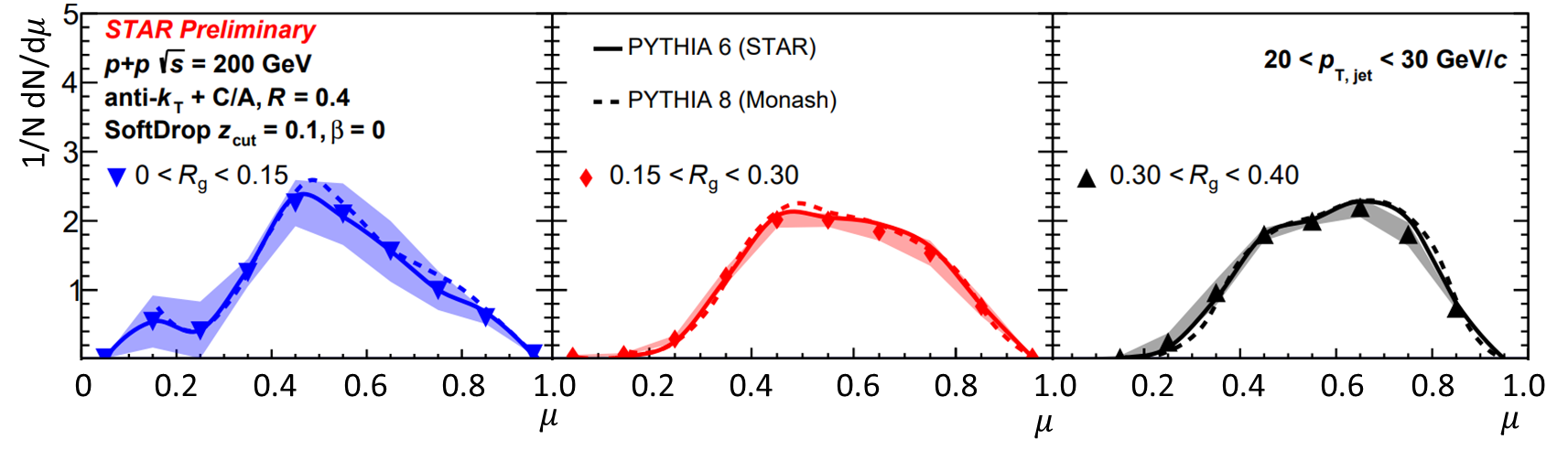}
    \caption{The $\mu$ distributions for various $R_{\mathrm{g}}$ selections.}
    \label{fig:mu}
\end{figure}

To provide further insight into the relation between the angular and mass scales of a jet, we turn to another new STAR result which reports the measurement of $\mu$ \cite{Dasgupta:2013ihk}, an observable that captures the mass sharing of the hard splitting and is defined as:

\begin{equation}
    \mu = \frac{\max(M_1,M_2)}{M_{\mathrm{g}}},
\end{equation}
where $M_{1,2}$ are the masses of the two subjets. Note that this measurement is corrected for detector effects with a (2+1)D Bayesian unfolding method \cite{Robotkova:2022jmk}, and does not require $M>1$ GeV$/c^2$, but imposes otherwise identical selection criteria as described in Sec. \ref{analysis_details}. Figure \ref{fig:mu} shows the distributions of $\mu$ for various ranges of $R_{\mathrm{g}}$. We observe that the trend in data is well described by PYTHIA6 STAR tune \cite{Adkins:2015ccl} and PYTHIA8 Monash tune \cite{Skands:2014pea}, and that $\mu$ has a weaker dependence on $R_{\mathrm{g}}$ compared to $\Delta M/M$. This measurement offers complementary information to the measurement in Fig. \ref{fig:dmfrac_rg}, revealing the correlation between the hard radiation and angle of the hard splitting. The shift of $\mu$ to smaller values at smaller $R_{\mathrm{g}}$ indicates that a narrower splitting leads to a smaller transfer of virtuality.

\section{Conclusions}
We have presented measurements of the CollinearDrop groomed jet mass and its correlation with the SoftDrop groomed observables $R_{\mathrm{g}}$ and $z_{\mathrm{g}}$, in $pp$ collisions at $\sqrt{s}=200$ GeV at STAR. MultiFold, a machine learning based unfolding method that enables access of multi-dimensional correlations on a jet-by-jet basis, has been shown to offer consistent results as RooUnfold in one-dimensional distributions, and has been applied to correct for the correlation measurements in higher dimensions. These measurements allow for a better understanding of the soft phase space within jets, and highlight how early-stage radiation affects the angular scale and momentum imbalance of a later splitting.

\printbibliography

@PREAMBLE{
 "\providecommand{\noopsort}[1]{}" 
 # "\providecommand{\singleletter}[1]{#1}%" 
}

@article{Dasgupta:2013ihk,
    author = "Dasgupta, Mrinal and Fregoso, Alessandro and Marzani, Simone and Salam, Gavin P.",
    title = "{Towards an understanding of jet substructure}",
    eprint = "1307.0007",
    archivePrefix = "arXiv",
    primaryClass = "hep-ph",
    reportNumber = "CERN-PH-TH-2013-145, DCPT-13-86, IPPP-13-43, LPN13-036, MAN-HEP-2013-12",
    doi = "10.1007/JHEP09(2013)029",
    journal = "JHEP",
    volume = "09",
    pages = "029",
    year = "2013"
}

@article{Brun:1994aa,
    author = "Brun, Ren\'e and Bruyant, F. and Carminati, Federico and Giani, Simone and Maire, M. and McPherson, A. and Patrick, G. and Urban, L.",
    title = "{GEANT} Detector Description and Simulation Tool",
    reportNumber = "CERN-W5013, CERN-W-5013, W5013, W-5013",
    doi = "10.17181/CERN.MUHF.DMJ1",
    month = "10",
    year = "1994"
}

@article{Skands:2010ak,
    author = "Skands, Peter Zeiler",
    title = "{Tuning Monte Carlo} Generators: The {Perugia} Tunes",
    eprint = "1005.3457",
    archivePrefix = "arXiv",
    primaryClass = "hep-ph",
    reportNumber = "MCNET-10-08, CERN-PH-TH-2010-113",
    doi = "10.1103/PhysRevD.82.074018",
    journal = "Phys. Rev. D",
    volume = "82",
    pages = "074018",
    year = "2010"
}

@phdthesis{Adkins:2015ccl,
    author = "Adkins, James Kevin",
    title = "Studying Transverse Momentum Dependent Distributions in Polarized Proton Collisions via Azimuthal Single Spin Asymmetries of Charged Pions in Jets",
    eprint = "1907.11233",
    archivePrefix = "arXiv",
    primaryClass = "hep-ex",
    school = "Kentucky U.",
    year = "2015",
    doi = "10.48550/arXiv.1907.11233"
}

@article{Sjostrand:2006za,
    author = "Sjostrand, Torbjorn and Mrenna, Stephen and Skands, Peter Z.",
    title = "{PYTHIA} 6.4 Physics and Manual",
    eprint = "hep-ph/0603175",
    archivePrefix = "arXiv",
    reportNumber = "FERMILAB-PUB-06-052-CD-T, LU-TP-06-13",
    doi = "10.1088/1126-6708/2006/05/026",
    journal = "JHEP",
    volume = "05",
    pages = "026",
    year = "2006"
}

@article{Skands:2014pea,
    author = "Skands, Peter and Carrazza, Stefano and Rojo, Juan",
    title = "{Tuning PYTHIA 8.1: the Monash 2013 tune}",
    eprint = "1404.5630",
    archivePrefix = "arXiv",
    primaryClass = "hep-ph",
    reportNumber = "CERN-PH-TH-2014-069, MCNET-14-08, OUTP-14-05P",
    doi = "10.1140/epjc/s10052-014-3024-y",
    journal = "Eur. Phys. J. C",
    volume = "74",
    number = "8",
    pages = "3024",
    year = "2014"
}

@article{cacciari2008anti,
    author = "Cacciari, Matteo and Salam, Gavin P. and Soyez, Gregory",
    title = "{The anti-$k_t$ jet clustering algorithm}",
    eprint = "0802.1189",
    archivePrefix = "arXiv",
    primaryClass = "hep-ph",
    reportNumber = "LPTHE-07-03",
    doi = "10.1088/1126-6708/2008/04/063",
    journal = "JHEP",
    volume = "04",
    pages = "063",
    year = "2008"
}

@article{Gribov:1972ri,
    author = "Gribov, V. N. and Lipatov, L. N.",
    title = "{Deep inelastic ep scattering in perturbation theory}",
    reportNumber = "IPTI-381-71",
    journal = "Sov. J. Nucl. Phys.",
    volume = "15",
    pages = "438--450",
    year = "1972"
}

@article{ALTARELLI1977298,
title = {Asymptotic freedom in parton language},
journal = {Nucl. Phys. B},
volume = {126},
number = {2},
pages = {298-318},
year = {1977},
issn = {0550-3213},
doi = {10.1016/0550-3213(77)90384-4},
author = {G. Altarelli and G. Parisi},
}

@article{Dokshitzer:1977sg,
    author = "Dokshitzer, Yuri L.",
    title = "Calculation of the Structure Functions for {Deep Inelastic Scattering} and $e^{+}e^{-}$ Annihilation by Perturbation Theory in {Quantum Chromodynamics}",
    journal = "Sov. Phys. JETP",
    volume = "46",
    pages = "641--653",
    year = "1977"
}

@article{larkoski2014soft,
    author = "Larkoski, Andrew J. and Marzani, Simone and Soyez, Gregory and Thaler, Jesse",
    title = "{Soft Drop}",
    eprint = "1402.2657",
    archivePrefix = "arXiv",
    primaryClass = "hep-ph",
    reportNumber = "MIT-CTP-4531, DCPT-14-24, IPPP-14-12",
    doi = "10.1007/JHEP05(2014)146",
    journal = "JHEP",
    volume = "05",
    pages = "146",
    year = "2014"
}

@article{Dokshitzer:1997in,
    author = "Dokshitzer, Yuri L. and Leder, G. D. and Moretti, S. and Webber, B. R.",
    title = "{Better jet clustering algorithms}",
    eprint = "hep-ph/9707323",
    archivePrefix = "arXiv",
    reportNumber = "CAVENDISH-HEP-97-06",
    doi = "10.1088/1126-6708/1997/08/001",
    journal = "JHEP",
    volume = "08",
    pages = "001",
    year = "1997"
}

@inproceedings{Wobisch:1998wt,
    author = "Wobisch, M. and Wengler, T.",
    title = "{Hadronization corrections to jet cross-sections in deep inelastic scattering}",
    booktitle = "{Workshop on Monte Carlo Generators for HERA Physics (Plenary Starting Meeting)}",
    eprint = "hep-ph/9907280",
    archivePrefix = "arXiv",
    reportNumber = "PITHA-99-16",
    pages = "270--279",
    year = "1998",
}

@book{Marzani_2019,
	doi = {10.1007/978-3-030-15709-8},
  
	url = {https://doi.org/10.1007%2F978-3-030-15709-8},
  
	year = 2019,
	publisher = {Springer International Publishing},
  
	author = {Simone Marzani and Gregory Soyez and Michael Spannowsky},
  
	title = {Looking Inside Jets}
}

@article{adam2020measurement,
  title={Measurement of groomed jet substructure observables in p+p collisions at $\sqrt{s}= 200$ {GeV} with {STAR}},
  collaboration={STAR},
  author={Adam, Jaroslav and others},
  journal={Phys. Lett. B},
  volume={811},
  pages={135846},
  year={2020},
  publisher={Elsevier},
  doi={10.1016/j.physletb.2020.135846}
}

@article{CMS:2017qlm,
    author = "Sirunyan, Albert M and others",
    collaboration = "CMS",
    title = "Measurement of the Splitting Function in $pp$ and {Pb-Pb} Collisions at {$\sqrt{s_{\mathrm{NN}}} =$ 5.02 TeV}",
    eprint = "1708.09429",
    archivePrefix = "arXiv",
    primaryClass = "nucl-ex",
    reportNumber = "CMS-HIN-16-006, CERN-EP-2017-205",
    doi = "10.1103/PhysRevLett.120.142302",
    journal = "Phys. Rev. Lett.",
    volume = "120",
    number = "14",
    pages = "142302",
    year = "2018"
}

@article{Aad:2705512,
    title = {Measurement of soft-drop jet observables in $pp$ collisions with the {ATLAS} detector at $\sqrt{s}=13\text{ }\mathrm{TeV}$},
author="Aad, G. and others",
collaboration = "ATLAS",
      journal = {Phys. Rev. D},
  volume = {101},
  issue = {5},
  pages = {052007},
  numpages = {37},
  year = {2020},
  publisher = {American Physical Society},
  doi = {10.1103/PhysRevD.101.052007},
  url = {https://link.aps.org/doi/10.1103/PhysRevD.101.052007}
}

@article{ATLAS:2017zda,
    author = "Aaboud, Morad and others",
    collaboration = "ATLAS",
    title = "{Measurement of the Soft-Drop jet mass in $pp$ collisions at $\sqrt{s} = 13$ TeV with the ATLAS Detector}",
    eprint = "1711.08341",
    archivePrefix = "arXiv",
    primaryClass = "hep-ex",
    reportNumber = "CERN-EP-2017-231",
    doi = "10.1103/PhysRevLett.121.092001",
    journal = "Phys. Rev. Lett.",
    volume = "121",
    number = "9",
    pages = "092001",
    year = "2018"
}

@article{chien2020collinear,
    author = "Chien, Yang-Ting and Stewart, Iain W.",
    title = "{Collinear Drop}",
    eprint = "1907.11107",
    archivePrefix = "arXiv",
    primaryClass = "hep-ph",
    reportNumber = "MIT-CTP 5034",
    doi = "10.1007/JHEP06(2020)064",
    journal = "JHEP",
    volume = "06",
    pages = "064",
    year = "2020"
}

@article{adam2020underlying,
  title={Underlying event measurements in $p$+$p$ collisions at $\sqrt{s}= 200$ {GeV at RHIC}},
  collaboration={STAR},
  author={Adam, Jaroslav and others},
  journal={Phys. Rev. D},
  volume={101},
  number={5},
  pages={052004},
  year={2020},
  publisher={APS},
  doi = {10.1103/PhysRevD.101.052004}
}

@article{andreassen2020omnifold,
  title={Omni{F}old: A method to simultaneously unfold all observables},
  author={Andreassen, Anders and Komiske, Patrick T and Metodiev, Eric M and Nachman, Benjamin and Thaler, Jesse},
  journal={Phys. Rev. Lett.},
  volume={124},
  number={18},
  pages={182001},
  year={2020},
  publisher={APS},
  doi={10.1103/PhysRevLett.124.182001},
}

@article{Robotkova:2022jmk,
    author = "Robotkov\'a, Monika",
    collaboration = "STAR",
    title = "{Multi-dimensional measurements of parton shower in $p+p$ collisions at RHIC}",
    doi = "10.22323/1.414.1183",
    journal = "PoS",
    volume = "ICHEP2022",
    pages = "1183",
    year = "2022"
}

@article{abdallah2021invariant,
  title={Invariant jet mass measurements in $pp$ collisions at $\sqrt{s}= 200\ \mathrm{GeV}$ at {RHIC}},
  author={Abdallah, MS and others},
  collaboration={STAR},
  journal={Phys. Rev. D},
  volume={104},
  number={5},
  pages={052007},
  year={2021},
  publisher={APS},
  doi={10.1103/PhysRevD.104.052007}
}

@article{aguilar2022pythia,
  title={{PYTHIA 8 underlying event tune for RHIC energies}},
  author={Aguilar, Manny Rosales and Chang, Zilong and Elayavalli, Raghav Kunnawalkam and Fatemi, Renee and He, Yang and Ji, Yuanjing and Kalinkin, Dmitry and Kelsey, Matthew and Mooney, Isaac and Verkest, Veronica},
  journal={Phys. Rev. D},
  volume={105},
  number={1},
  pages={016011},
  year={2022},
  publisher={APS},
  doi = {10.1103/PhysRevD.105.016011}
}

@article{bellm2016herwig,
  title={Herwig 7.0/{H}erwig++ 3.0 release note},
  author={Bellm, Johannes and others},
  journal={Eur. Phys. J. C},
  volume={76},
  number={4},
  pages={1--8},
  year={2016},
  publisher={Springer},
  doi={10.1140/epjc/s10052-016-4018-8}
}

@article{d2010improved,
  title={Improved iterative {B}ayesian unfolding},
  author={D'Agostini, Giulio},
  doi={10.48550/arxiv.1010.0632},
  year={2010}
}

@article{STAR:2002eio,
    author = "Ackermann, K. H. and others",
    collaboration = "STAR",
    title = "{STAR detector overview}",
    doi = "10.1016/S0168-9002(02)01960-5",
    journal = "Nucl. Instrum. Meth. A",
    volume = "499",
    pages = "624--632",
    year = "2003"
}

@article{CMS:2018fof,
    author = "Sirunyan, Albert M and others",
    collaboration = "CMS",
    title = "{Measurement of the groomed jet mass in PbPb and pp collisions at $ \sqrt{s_{\mathrm{NN}}}=5.02 $ TeV}",
    eprint = "1805.05145",
    archivePrefix = "arXiv",
    primaryClass = "hep-ex",
    reportNumber = "CMS-HIN-16-024, CERN-EP-2018-097",
    doi = "10.1007/JHEP10(2018)161",
    journal = "JHEP",
    volume = "10",
    pages = "161",
    year = "2018"
}

@article{Larkoski:2017jix,
    author = "Larkoski, Andrew J. and Moult, Ian and Nachman, Benjamin",
    title = "Jet Substructure at the {Large Hadron Collider}: A Review of Recent Advances in Theory and Machine Learning",
    eprint = "1709.04464",
    archivePrefix = "arXiv",
    primaryClass = "hep-ph",
    doi = "10.1016/j.physrep.2019.11.001",
    journal = "Phys. Rept.",
    volume = "841",
    pages = "1--63",
    year = "2020"
}

@book{Dokshitzer:1991wu,
    author = "Dokshitzer, Yuri L. and Khoze, Valery A. and Mueller, Alfred H. and Troian, S. I.",
    title = "{Basics of perturbative QCD}",
    year = "1991"
}

\end{document}